\documentstyle[aps,pre,multicol,epsf]{revtex}

\begin{document}

\title{Barkhausen avalanches in anisotropic ferromagnets with
  $180^\circ$ domain walls}

\author{Bosiljka Tadi\'c$^{1,*}$ and Ulrich Nowak$^{2,**}$}

\address{ $^1$Jo\v{z}ef Stefan Institute, P.O. Box 3000, 1001
  Ljubljana, Slovenia $^2$Theoretische Tieftemperaturphysik,
  Gerhard-Mercator-Universit\"at-Duisburg, D-47048 Duisburg, Germany }


\maketitle
\begin{abstract}
  We show that Barkhausen noise in two-dimensional disordered
  ferromagnets with extended domain walls is characterized by the
  avalanche size exponent $\tau _s =1.54$ at low disorder.  With
  increasing disorder the characteristic domain size is reduced
  relative to the system size due to nucleation of new domains and a
  dynamic phase transition occurs to the scaling behavior with $\tau
  _s=1.30$.  The exponents decrease at finite driving rate.  The
  results agree with recently observed behavior in amorphous Metglas
  and Fe-Co-B ribbons when the applied anisotropic stress is varied.
  \end{abstract}
\pacs{PACS numbers: 05.65.+b, 75.60.Ej, 68.35.Ct,05.40.-a }

\begin{multicols}{2}
Barkhausen noise measured in disordered ferromagnets at low
temperatures under the condition of slow driving by an external
field through the hysteresis loop exhibits scaling behavior without
tuning of any parameter
\cite{MC1,Torino,Urbach,BGD,D+Bs,Durin-stress,Brasil}.  The
scaling properties of Barkhausen avalanches obtained in various
alloys can be grouped in three distinct universality classes
differentiated by the value of the avalanche size exponent as
\cite{BT-Calcutta} $\tau _s\approx 1.3$, $\tau _s\approx 1.5$, and
$\tau_s \approx 1.7$. However, the origin of the scaling of
Barkhausen noise (BN) and the occurrence of different universality
classes is still not fully understood.

The above mentioned measurements
\cite{MC1,Torino,Urbach,BGD,D+Bs,Durin-stress,Brasil} are done
mostly on thin ribbon samples (thickness $d < 200 \mu m$).
The variety of
measured scaling exponents can be attributed to the differences in the
applied driving conditions and to the diversity of domain structures
occurring in different samples.   It was found that the exponents
decrease continuously with increasing driving rate
\cite{Torino,D+Bs,Durin-stress}.  The structure of
domains in commercial alloys has been studied by variety of techniques
(see for instance \cite{LM}). In thin ribbons of the amorphous Metglas
Fe-B-Si \cite{MC1,Brasil,LM} and Fe-Co-B alloys
\cite{D+Bs,Durin-stress} which are annealed in a parallel field
\cite{LM} or under an applied anisotropic stress
\cite{Durin-stress,Brasil} a structure with few domains occurs
with $180^{\circ}$ domain walls parallel to the anisotropy axis. The
demagnetizing fields, which depend on the form of the sample, and the
range of interactions play an important role for the equilibrium
domain structure as well as for the domain-wall dynamics
\cite{Urbach,D+Bs}.  It has been demonstrated that the domain
structure in the amorphous Metglass \cite{Brasil} and Fe-Co-B alloys
\cite{Durin-stress} can be controlled by varying tensile stress and
that short-range interactions dominate over dipolar forces
\cite{Durin-stress}.  In addition, in these systems the demagnetizing
fields are minimized with longitudinal anisotropy \cite{LM}. Hence, in
our study we can neglect dipolar forces, but we take into account the
existence of domain walls.  Numerical studies of BN using short-range
Ising models with various types of disorder \cite{RF,RB,BT} (see also
\cite{Gonsales} for more realistic type of interactions) usually start
from a uniform ground state and nucleate a random pattern of clusters
of reversed spins by increasing the external field, thus
neglecting  the preexisting domain structure.

In this paper we simulate Barkhausen avalanches using a model with
preexisting {\it extended} domain walls which are confined in two
dimensions, motivated by the stress-induced anisotropy in realistic
systems \cite{Durin-stress,Brasil}.  Increasing the external magnetic
field the domain wall may either move through a random medium or new
domains may nucleate when this is energetically favorable
\cite{Weller}. It has been suggested recently that interface depinning
in a random medium is responsible for the scaling behavior of BN
\cite{Urbach,D+Bs,Brasil}, however, models of an elastic interface
yield scaling exponents which are lower than the measured ones
\cite{Urbach,Brasil}. In this work we use a ferromagnetic model with
short-range interactions and a random-field pinning and we show that
the results compare well with two universality classes of measured
scaling exponents. In addition, a dynamic phase transition between
these two scaling behaviors appears when the strength of disorder (or
size of the domains) is tuned.  It has been recognized recently that
disorder effects are enhanced leading to smaller domain sizes by
decreasing either tensile stress \cite{Durin-stress,Brasil} or grain
sizes \cite{Gonsales}.  Motivated by these suggestions, here we study
the influence of disorder on the scaling properties of Barkhausen
noise.  We expect that the results are
relevant to realistic samples in which
the domain size exceeds the sample thickness.

We consider an Ising model on a square lattice of size $L \times L$
 assuming that local random fields $h_i$ are
generated by coarse-graining from an original disorder \cite{Weller}
in the presence of the external magnetic field $H$:
\begin{equation}
  {\cal{H}} = -\sum_{<i,j>}J_{i,j}S_iS_j - \sum_i (h_i+H)S_i \ .
  \label{Hamiltonian}
\end{equation}
Here we set $J_{i,j}=1$ to be a constant interaction between
nearest-neighbor spins $S_i = \pm 1$. Hence, all fields and energies are
measured in units of $J_{i,j}$. A Gaussian distribution of $h_i$ is
assumed with zero mean and width $f$.  We create an initial domain
wall in $\langle 11 \rangle$ direction by a rotation of the lattice by
$\pi /4$ and setting all spins except of those in the first row
opposite to the external field \cite{Uli-T}.  Periodic boundaries in
the direction of the interface and fixed boundaries in the
perpendicular direction are applied.  The dynamics consists of a spin
alignment parallel to the external field when the local field
$h_i^{loc} \equiv \sum_jJ_{i,j}S_j +H+h_i$ exceeds zero.  For the
simulation of a hysteresis loop the field updates are adjusted to the
minimum local field (infinitely slow driving), thus the driving field
is uniform in space but fluctuates in time.  We also briefly discuss
the effects of finite driving rates.

It should be stressed that at each time step we update {\it all
  spins}, which is suitable for a globally driven magnetic system.  In
this way new domains are nucleated when it is energetically allowed,
in contrast to models of driven interfaces, where an update is
restricted only to the sites located next to the
interface, resulting in ``percolation'' growth even
at high disorder \cite{JR}.  Another important feature of our model is
the anisotropy \cite{TKD} due to the initial conditions:
the threshold driving forces in the parallel and
perpendicular directions appear to be different.  This model has
the following  advantages:  (1) {\it In the limit of vanishing disorder
the $\langle 11 \rangle$ interface can be moved by an infinitesimally
small  field}.
This bypasses the problem of threshold energy $2J$, which occurs in an
$\langle 10 \rangle $ interface \cite{JR} implying that a large
lattice size have to be used in order to find a spin with a random
field large enough to surmount the energy barrier. Therefore, here we
can apply smaller lattice sizes and  vary disorder configurations
(we use up to $2\times 10^3$ configurations and up to $L=400$).
(2) The $\langle 11 \rangle$ interface depinns at infinitesimally
small field at low disorder $f\to 0$. Hence, an {\it
upper limit of disorder $f^\star $ exists} at which depinning is no
longer possible, and nucleation of new domains in the interior
becomes favorable at large enough fields. This feature comprises an
important difference compared to the model of Ref.\ \cite{RF}, where
 nucleation of a single spanning cluster at $f\to 0$ requires energy
threshold $4J$, and thus becomes obscured by finite lattice size and
large fields.

Fig.\ 1 shows snapshots of simulated systems for high and low
disorder. For low disorder only domain wall motion occurs.
 For values
of $f$ that exceed a certain critical value $f^\star$ (determined
below), domains nucleate inside the system, thus leading to the same
structure of clusters as in systems without an initial domain wall
\cite{BT,TN}.  These two regions are shown in the phase diagram in
Fig.\ 2 together with simulation results for the coercive fields
$H_0(f)$ of the hysteresis and the critical fields $H_c(f)$ of the
depinning transition (explained below).

We apply the quasi-static driving described above (an example of time
series of field increments is given in the inset to Fig.\ 2) and
monitor the motion of domain walls.  The number of flipped spins $s$
between two consecutive locally stable configurations of the wall
determines the size of Barkhausen avalanche. The avalanche size
distribution $D(s,f)$ is shown in Fig.\ 3 for various values of disorder
$f$. Taking the avalanche statistics along the ascending part of the
hysteresis loop (until eventually depinning occurs) and averaging over
many disorder configurations we find the slopes according to $D(s)\sim
s^{-\tau _s}$ as $\tau _s =1.54$ for $f\le 0.6$, and $\tau _s=1.30$
above $f^\star \approx 0.6$. The cut-off also decreases with $f$.  A
similar behavior was found experimentally in Metglas 2605TCA in Refs.\
\cite{Brasil}. With the stress $\sigma $ varying in the range from 0
-- 525 MPa the slopes of the size distributions are reported to vary
from 1.29 to 1.60, and the cut-offs increase \cite{Brasil}.  In ${\rm
  Fe_{64}Co_{21}B_{15}}$ the measured slope was 1.28
\cite{Durin-stress} for $\sigma $ up to 140 MPa.  The applied
anisotropic stress stretches the domain walls, thus for the degree of
disorder $f$ we have $f\sim 1/\sigma ^x$.
We also measure the distributions of the linear
extensions of avalanches in the directions parallel ($w$) and
perpendicular ($h$) to the wall leading to the anisotropy exponent
$\zeta \equiv (\tau _w -1)/(\tau _h -1) =0.92$. (See Table\ 1).

\narrowtext
\begin{figure}
  \epsfxsize=80mm\epsffile[21 255 589 537]{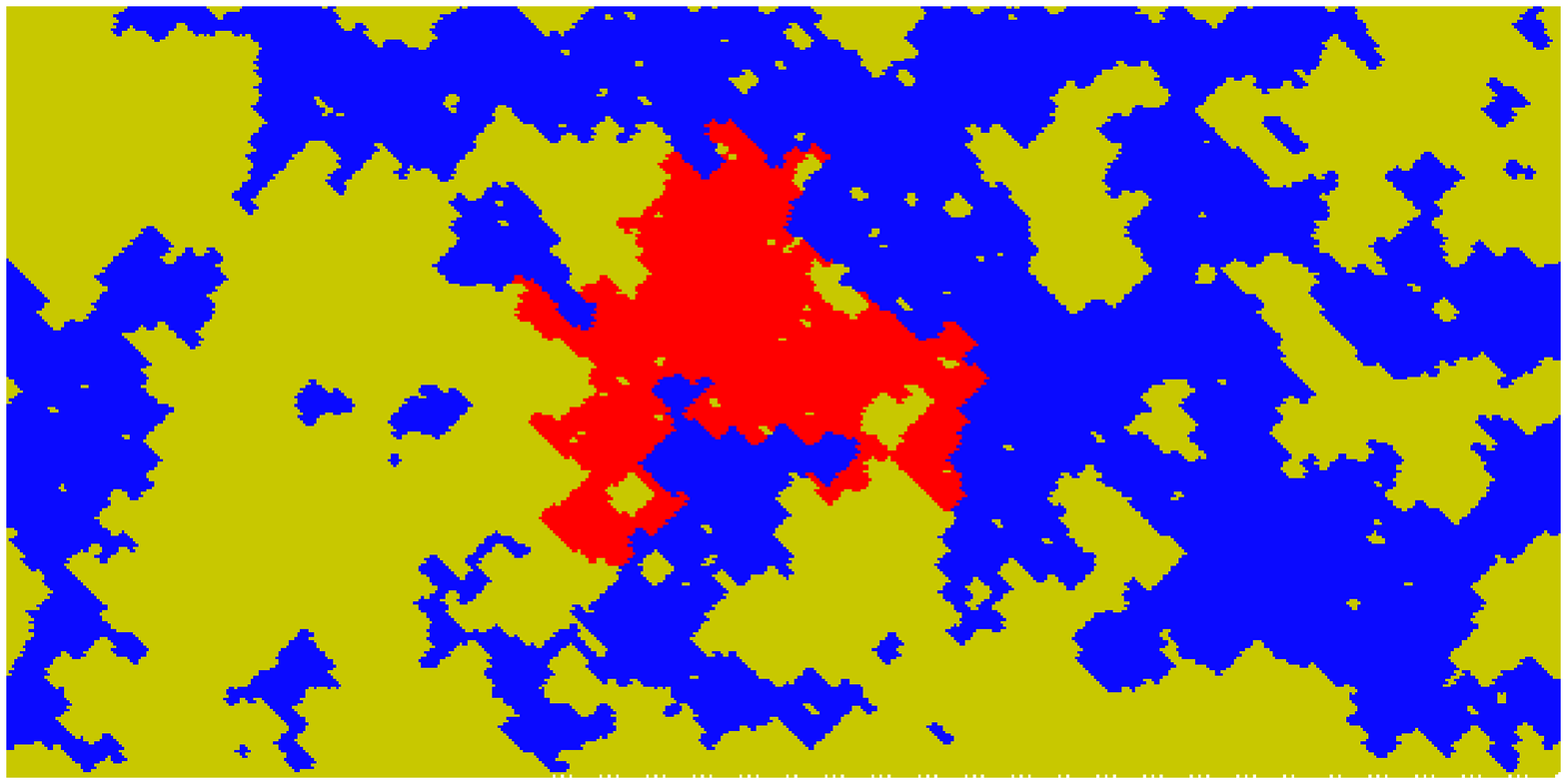}
  \epsfxsize=80mm\epsffile[21 255 589 537]{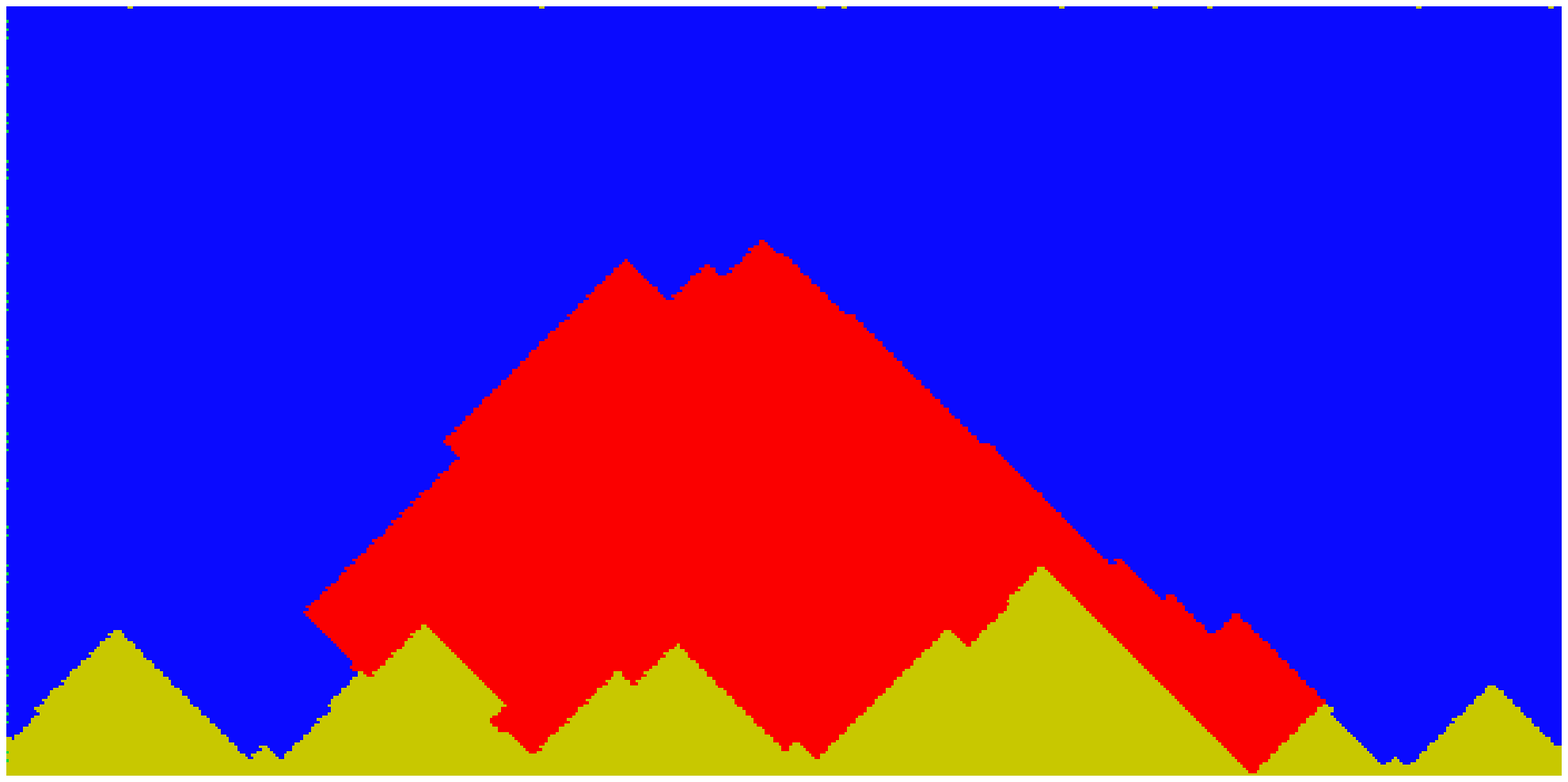}
  \caption {Emerging  cluster structure of flipped (bright) and
  unflipped (dark) spins, and most recent cluster (gray),
  illustrating self-affine growth at low disorder (bottom), and
nucleation and blocking by previous clusters at high disorder
(top) $L=200$.
}
  \label{fig1}
\end{figure}

In the low disorder region the domain wall depinns when the driving
field exceeds a critical value $H_c(f)$.  To investigate the depinning
transition we start with a flat wall and apply a constant field $H$
measuring the {\it long-time limit} of the domain wall velocity $v$,
which is defined as the number of flipped spins {\it per internal time
  step} relative to $L$.  The velocity averaged over  2000
disorder realizations, $[v]$, is  the order parameter of
the depinning transition. Here $[v]$ and its fluctuations are obtained
for $L=$100, 200, and 400 and the critical exponents and critical fields
$H_c(f)$ (also shown in Fig.\ 2) are determined by a finite size
scaling plot of the general form
\begin{equation}
  Y(X,L) = L^{-Z_Y}{\cal{Y}}(L^{1/\nu}X)
\label{FSS-H}
\end{equation}
with $X\equiv (H-H_c(f))/H_c(f)$ (see Table\ 1, left side, and Ref.\
\cite{TN}). Here $Z_Y\equiv \beta /\nu$  for
the analysis of the order parameter $[v]$, and $Z_Y\equiv\gamma /\nu$
for the analysis of the fluctuations of $[v]$. Our results
are compatible with the correlation length exponent $\nu =1.23 \pm
0.04$, $\beta =0.43 \pm 0.03$, and $\gamma =1.52 \pm 0.06$.  The
exponents are universal in the region $0.35 \le f \le 0.6$.  Using the
scaling relation $\beta /\nu =z-\zeta $ valid for the depinning
transition \cite{TKD} we find the dynamic exponent $z=$1.27.  Then the
relation $D(\tau _s-1)=z(\tau_t-1)$, where the fractal dimension of
anisotropic avalanches is $D=1+\zeta $, leads to the duration
exponent $\tau _t =$1.83. These results are in good agreement with
universal criticality in a class of driven dynamical systems recently
discussed in Ref.\ \cite{extremal}.  The fraction of
active sites at time $t$ scales as $N(t)\sim t^\kappa $, with $\kappa
=D/z-1 =0.51$, compared with 0.58 in \cite{Sneppen}.
The value for the exponent $\nu $ compares well with one in \cite{JR,Boston}.
For the (local)  roughness exponent $\zeta $ values in the
literature ranging from 0.5 to 1.23 can be attributed to the influence
of the anisotropy \cite{TKD}, elasticity of the interface
\cite{elastic}, distribution of disorder \cite{Uli-T,JR}, and
 driving conditions \cite{extremal,SOD}.
 Further  analysis, e.g., by the dynamic renormalization group, is
necessary in order to determine if the value $\zeta =0.92$ found in
this work represents a new universal behavior or a crossover due
to finite size effects.
This value suggests closeness to the class of models with  vanishing
interface velocity \cite{Boston} and  self-organized depinning \cite{SOD}.
At low disorder lateral motion of the interface makes it possible
to overcome strong pinning centers and to maintain the
self-affine growth (see Fig.\ 1). However, at the transition we find
 $\phi \equiv 2\nu(1-\zeta )-\beta =-0.23$, thus
$\lambda _{eff} \to 0$ resulting in a different critical  behavior
compared to the case $\lambda _{eff}\to \infty$ studied in Refs.\
 \cite{TKD,Boston}.

In order to determine the largest disorder where depinning is still
possible, $f^\star$, we can formally extend the above study of the
domain wall velocity now averaged over hysteresis loop and disorder
$[<v>]$, and the corresponding susceptibility.  Applying then Eq.\
(\ref{FSS-H}) with $X\equiv (f-f^\star )/f^\star$ (scaling collapse is
shown in the inset to Fig.\ 3) we find $f^\star =0.61\pm 0.02$ and the
exponents $\beta$, $\gamma$ and $\nu$ shown in Table \ 1, right side.

In the region of high disorder $f>f^\star $ nucleation of new domains
of finite size which are blocking each other's spatial extent becomes
the dominant feature which determines the scaling properties of BN
(see Table\ 1 (right)). The section of power-law behavior of the
distribution of avalanches increases with decreasing $f$, in
qualitative agreement with increasing stress in experiments
\cite{Durin-stress,Brasil}. In the inset to Fig.\ 4 the distribution
of avalanche durations is shown for various values of $f$ above
$f^\star$.  The main Fig.\ 4 shows the scaling plot $P(t,f) = (\delta
f)^{z\nu \tau _t}{\cal{P}}(t(\delta f)^{z\nu })$, where $\delta f
\equiv (f-f^\star )/f^\star $, obtained by using $f^\star =0.62$ and
the products of the exponents $z\nu \tau _t=4.43$ and $z\nu =2.98 $.
>From a similar scaling collapse of the size distribution we find
$D\nu \tau _s =5.61$ and $D\nu =4.30$.  Thus,
these values together with the ones in Table\ 1 (right) lead to  $\nu =
2.3\pm 0.1$, $\beta =0.12\pm 0.03$, and $\gamma = 5.2\pm 0.1$.
Note that the value of
$f^\star = 0.62$ within statistical error bars $\pm 0.03$,
which we estimated from two different types of scaling fits
(see Figs.\ 3 and 4) is by no means definitive. (The  error bars are
expected to increase when more scaled quantities or wider range of
system sizes are explored). However, since no exact results are available,
this value can be regarded as a rough estimate of the critical disorder.

By applying a finite driving rate $r\equiv \Delta H/H_{max}$, where
$H_{max}$ is the saturation field, the cut-offs of the distributions
in the high disorder region increase (see also \cite{BT}).  Whereas
slopes of the distribution decrease with $r$
due to mainly the coalescence of avalanches.  For the size of
avalanches we find, for instance for $f=0.88$, $\tau _s = 1.18$
 for $r= 0.01$, and $\tau _s = 1.10$ for $r= 0.02$.

In conclusion, we have shown that the anisotropic 2-dimensional motion
of domain walls pinned by quenched impurities and short-range
interactions are relevant for the scaling behavior of Barkhausen noise
in the presence of extended domain walls.  We find two universality classes
with $\tau _s=1.54$ and $\tau _s=1.30$, in a good agreement with
experiments in amorphous Fe-B-Si and Fe-Co-B ribbons under anisotropic
stress.  By slow driving the scaling behavior is robust in a wide
range of the driving field and disorder (or stress) values.  These two
universality classes correspond to the motion of extended domain walls
(i.e., at low disorder or high stress), and many finite domains (high
disorder or low stress), respectively.  The scaling exponents decrease
with finite driving rate. We find that the domain structure changes
via a dynamic phase transition at a critical disorder, which can be
directly monitored in experiments by tuning uniaxial stress.

\acknowledgments This work was supported by the Ministry of Science
and Technology of the Republic of Slovenia and by the Deutsche
Forschungsgemeinschaft through Sonderforschungsbereich 166.


\narrowtext
\begin{table}
  \begin{center}
    \begin{tabular}{|c|c||c|c|}
      Exponent& Value (LD) &Exponent &Value (HD)\\ \hline $\tau _s$&
      1.54&$\tau _s $& 1.30 \\ $\tau _w$& 2.03&$\tau _t$& 1.47\\ $\tau
      _h$&2.12& $D$&1.88\\ $\zeta $ &0.92&$z $&1.23\\ \hline $\beta
      /\nu $&0.35& $\beta /\nu$&$0.06$\\ $\gamma /\nu $&1.22& $\gamma
      /\nu$ &2.26\\ $1/\nu $ &0.82&$1/\nu$ & 0.42\\
    \end{tabular}
  \caption{Scaling exponents obtained by {\it direct} simulation of
    various quantities for the low (LD) and high (HD) disorder region.
    Error bars estimated from independent fits are within $\pm 0.03$.}
  \label{table1}
  \end{center}

\narrowtext

\begin{figure}
  \epsfxsize=80mm\epsffile[48 68 581 579]{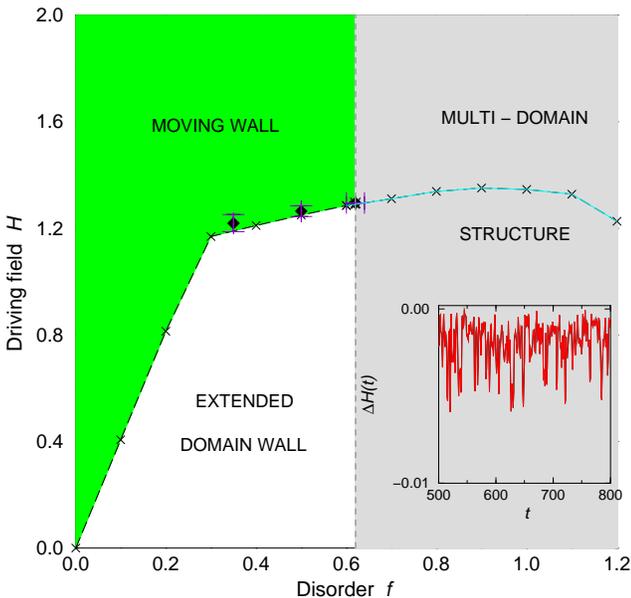} \vskip 2mm
\caption{Phase diagram in the plain disorder--driving-field
  (both in units of $J$). The vertical dashed line at $f^\star =0.62$
  separates high and low-disorder region.  Long-dashed line with
  crosses: coercive fields $H_0(f)$; filled symbols: critical fields
  $H_c(f)$ obtained from scaling fits. Inset: Time (in MCS) series of
  driving field increments $\Delta H$ (in J/MCS) for $f=1.2$.}
\label{fig2}
\end{figure}

\begin{figure}
  \epsfxsize=80mm\epsffile[38 70 522 540]{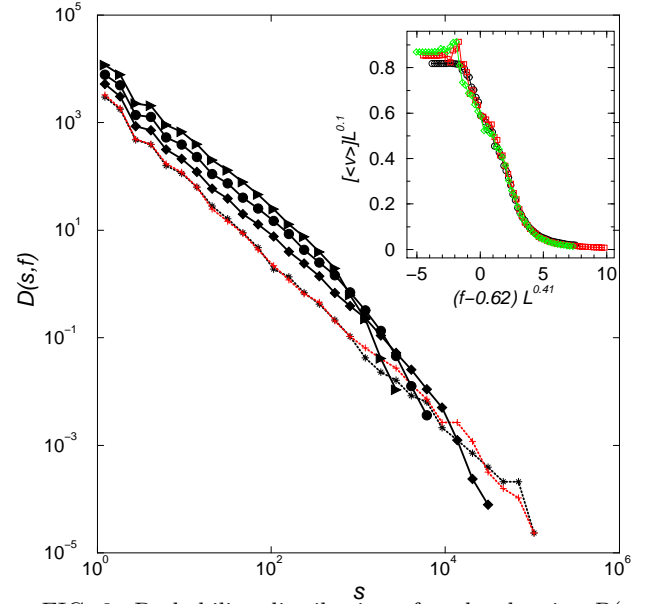}
\caption{Probability distribution of avalanche size $D(s,f)$ vs size $s$
  for various values of disorder $f$=0.4, 0.6, 0.9, 1.0, and 1.1
  (bottom to top).  Inset: Scaling collapse according to Eq.\
  (\ref{FSS-H}) of the velocity of domain wall $[<v>]$ vs. $f$ for
  $L$=128, 196, and 256. Each point is averaged over 200 samples.}
\label{fig3}
\end{figure}

\begin{figure}
  \epsfxsize=80mm\epsffile[36 68 582 580]{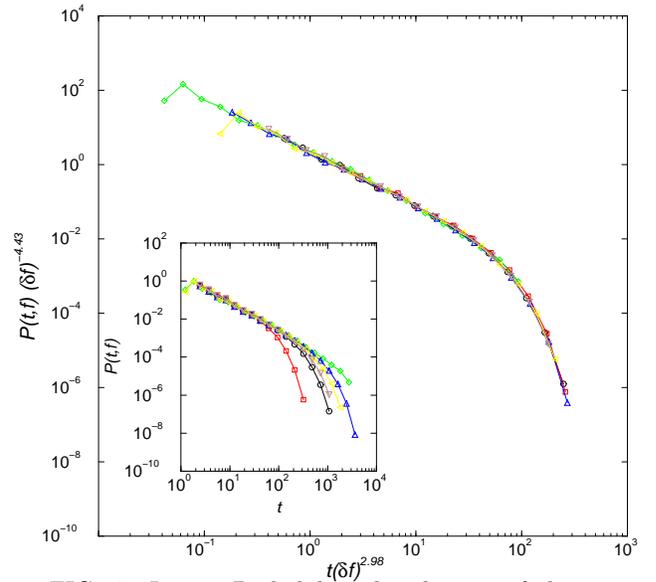}
\caption{Inset: Probability distribution of durations  of
  Barkhausen avalanches $P(t,f)$ vs. duration $t$ in MCS, for $L=400$
  and for $f$= 1.2, 1.0, 0.96, 0.92, 0.88, and 0.84 (left to right).
  Main figure: Scaling collapse of the data (see text). }
\label{fig4}
\end{figure}
\end{table}

\end{multicols}

\end{document}